**Parallel Core-Shell Metal-Dielectric-Semiconductor Germanium Nanowires for High Current Surround Gate Field Effect Transistors**

Li Zhang, Ryan Tu and Hongjie Dai*

Department of Chemistry and Laboratory for Advanced Materials, Stanford University, Stanford, CA 94305, USA

**Abstract**

Core-shell germanium nanowire (GeNW) is formed with a single-crystalline Ge core and concentric shells of nitride and silicon passivation layer by chemical vapor deposition (CVD), an $Al_2O_3$ gate dielectric layer by atomic layer deposition (ALD) and an Al metal surround-gate (SG) shell by isotropic magnetron sputter deposition. Surround gate nanowire field effect transistors (FETs) are then constructed using a novel self-aligned fabrication approach. Individual SG GeNW FETs show improved switching over GeNW FETs with planar gate stacks owing to improved electrostatics. FET devices comprised of multiple quasi-aligned SG GeNWs in parallel are also constructed. Collectively, tens of SG GeNWs afford on-currents exceeding 0.1mA at low source-drain bias voltages. The self-aligned surround gate scheme can be generalized to various semiconductor nanowire materials.

* Email: hdai@stanford.edu

GeNWs have attracted much attention as building blocks for future nanoelectronic components owing to their low temperature synthesis and high bulk mobility.[1-9] An active area of research has been the continual optimization of FETs based on individual NWs.[5, 8, 10] Such devices are typically fabricated in the plane of a substrate with either a top or bottom gate. It is well known that a surround gate structure, whereby the gate fully wraps around the channel, is optimal for electrostatic control over charge carriers in the channel.[11] Chemically synthesized NWs offer an advantage over top-down lithographically patterned semiconductor wafers for the realization of SG FETs. Vertical SG NW FETs have already been demonstrated using epitaxially grown NWs, although the fabrication generally requires multiple complex steps and high temperatures.[12-15] Another area of research is the fabrication of FETs with multiple, parallel NWs in each FET in order to reach sufficiently high on-currents to drive practical circuits.[16, 17]

Here, we present SG GeNW FETs based on individual and parallel arrays of core-shell metal-dielectric-semiconductor GeNWs, with on-current exceeding 0.1mA for the latter. The cylindrical GeNW is fully surrounded by a concentric shell of $Al_2O_3$ gate dielectric and Al gate metal for optimum electrostatic control of the channel. A self-aligned fabrication process is developed to minimize the un-gated length of NWs and parasitic capacitance. The wrapped around geometry improves on/off ratios and sub-threshold swings of GeNW FETs with planar gate stacks. Our fabrication process is simple and can be generalized to obtain SG FETs of various types of semiconducting NWs, especially for those that are difficult to grow epitaxially on substrates or require low thermal budget processes. For multiple-wire FETs, the use of SG NWs is advantageous since each wire has its own surrounding gate shells. Electrostatic shielding and interference by neighboring or crossing NWs is avoided or minimized.

GeNWs were synthesized by CVD of $GeH_4$ at 295°C on Au nanocolloids (20nm in diameter) densely dispersed on $SiO_2$ substrates.[2, 3] As grown NWs formed a forest and were observed by cross sectional SEM to be standing out of the substrate with most NWs pointing within 30° of the plane normal (Fig. 1a). The wires were *in-situ* annealed in 10% $NH_3$ in Ar followed by 1.99% $SiH_4$ in Ar at 400°C to afford a thin passivation layer of nitride and silicon (Fig.1b, step i, thickness ~ 1.25nm ).[18, 19] Only the first monolayer of Ge is nitrided for low temperature $NH_3$ annealing below 600°C.[20] We then deposited 4 nm of $Al_2O_3$ conformally around the GeNWs (Fig.1b, step ii) by ALD[21, 22] in a separate reactor at 100 °C using a precursor of trimethyl aluminum (TMA) followed by 15nm of Al by nearly isotropic magnetron sputter deposition (Fig.1b, step iii). The Si overlayer was oxidized by ambient air to form $SiO_x$ when exposed to air during transferring to the ALD reactor.  Due to the nearly free-standing nature of as-grown GeNWs (Fig.1a), isotropic and conformal dielectric ALD, and non-directional metal deposition by sputtering, our process afforded core-shell $Al/Al_2O_3/Ge$ NWs with approximate cylindrical geometry, as confirmed by transmission electron microscopy (TEM, Fig.1b).

The core-shell GeNWs were then sonicated off the substrate in isopropanol alcohol (IPA) to afford a suspension. For fabrication of SG FETs of individual NW, droplets of the suspension were spin-coated onto a Si substrate with 500nm of thermally grown $SiO_2$. Lithographic patterning was used to open windows in polymethyl methacrylate (PMMA) over source (S) and drain (D) regions (Fig.2a, step i) of a nanowire and define a ~3μm channel length. The Al and $Al_2O_3$ shells on a GeNW in the opened PMMA windows were etched for 4 minutes by a dilute solution of 0.01M KOH in 95% $H_2O$ and 5% IPA (Fig.2a, step ii). Since the wet etching is isotropic, the Al and $Al_2O_3$ on the GeNWs were undercut at the PMMA edges of the opened windows. Directional electron-beam evaporation of 60nm Ti followed by liftoff was used to complete the S/D contacts. A second patterning step was then carried out to contact the outer Al shell of the SG GeNW by a narrow Pt electrode (Fig. 2a, step iii) to complete the gate connection. Lastly, the sample was annealed in forming gas at 300°C for 30 minutes to improve the contacts between the S/D and the GeNW.

The undercutting process during KOH etching of Al and $Al_2O_3$ shells in the source and drain regions was important to preventing the deposited S/D metal from shorting to the SG metal, and affording self-aligned S/D and SG, with a small gap (~40nm due to undercutting, visible in the inset of Fig. 2c) between the edges of S/D contacts and the surround metal gate shell. The GeNW in the gap remained passivated by $SiO_x$ due to its low etch rate by dilute KOH (<0.1A/min).[23] The use of $Al/Al_2O_3$ shells and isotropic KOH etching can be generalized to the fabrication of self-aligned SG FETs for various semiconductor NWs. The relative ease of KOH etching of $Al_2O_3$ makes $Al_2O_3$ an advantageous dielectric material for SG NW FETs using our process. Other high $\kappa$ dielectrics such as $HfO_2$ and $ZrO_2$ tend to be more difficult to etch.

The electrical properties of our SG GeNW FETs (Fig.3a) exhibit p-type characteristics (due to light, unintentional p-doping in our growth system) with an on/off current ratio ($I_{on}/I_{off}$) of ~$10^5$ at -0.1V source-drain bias ($V_{ds}$) and a sub-threshold slope (SS) of 120mV/decade (Fig. 3b). The transconductance ($g_m$) at $V_{ds} = -0.1V$ is 0.33μS. It is known that the GeNW without passivation quickly forms an unstable oxide at the surface and the $Ge/GeO_2$ interface has been shown to introduce a high density of surface states.[24] Significant hysteresis during a double sweep of the unpassivated GeNW devices is caused by these surface states.[25] In contrast, a double sweep of our passivated SG structure shows no appreciable hysteresis (Fig. 3b inset). This suggests that the nitride and silicon passivation layer prevents oxidation of the GeNW surface. In addition, the SS and $I_{on}/I_{off}$ are significantly improved over our earlier results obtained with GeNW FETs with planar topgate stacks (see ref. 8 where SS~300mV/decade typically). These indeed suggest better switching characteristics of SG GeNW FETs. Current-gate voltage ($I_{ds}$-$V_{gs}$) transfer characteristics recorded at various biases up to -1V (Fig. 3b) show similar SS at high $V_{ds}$ as low biases, further suggesting good electrostatic control over the GeNW channel by the SG. $I_{ds}$-$V_{ds}$ curves at various gate biases (Fig. 3c) show a saturation on-current of ~ 4$\mu$A for a typical SG GeNW FET.

We estimated that the hole mobility ($\mu$) in our SG GeNW is ~ 197cm$^2$/Vs, calculated using the square law charge control model[26] at low bias $g_m$:

$$\mu = \frac{g_m L^2}{V_{ds} C} \qquad (1)$$

where L = 3$\mu$m is the channel length and C ~ 1.54fF is the gate capacitance calculated using a 2-D finite element electrostatic simulator (Estat 6.0, *Field Precision Software*) with geometry and thicknesses identical to our SG GeNWs (Fig. 1b). We used dielectric constants ($\varepsilon_0$) of 1.7 for the $SiO_x$ layer (~1.25nm thick) and 7.3 for the $Al_2O_3$ layer (~4nm thick), which were determined by direct capacitance-voltage measurements of planar Ge-$SiO_x$-$Al_2O_3$ stacks.[27]

Our SG mobility is lower than the best reported mobility[5] in GeNW FETs of 730cm$^2$/Vs and can be attributed to several factors. First, square law model assumes a transparent ohmic S/D contact where the current is not limited by the contact resistance. Our SG GeNW FETs have significant contact resistance due to Ti-Ge Schottky barriers and about 40nm ungated region near the S/D edges. Our contacts are not ohmic without heavy doping of the NWs in the source and drain regions like in a metal-oxide-semiconductor FET (MOSFET). Our work here focuses on developing the SG aspect of nanowire FETs without optimization of other elements such as doping and contacts. Second, the SG GeNW may still have significantly high density of interface states with an amorphous SiO$_x$ passivation layer. The combination of interface states and small bandgap of Ge may explain the high off-current. The best reported mobility was obtained for GeNW FETs when 1.7nm of crystalline Si was epitaxially grown around a GeNW core.[5, 10] Heteroepitaxially deposited Si could better passivate the GeNWs and minimize interfacial roughness. In addition, the valence band offset of an epitaxially grown crystalline Si shell affords ohmic contacts by shifting the Fermi level in the Ge core below the valence band.[10] Further improved performances and electrostatic control are expected when integrating SG structures into epitaxial Si/Ge NW FETs.[28]

Next, we fabricated GeNW FETs with multiple SG NWs in parallel in each transistor (Fig.4a). GeNWs were deposited onto a Si substrate with 500nm of thermally grown SiO$_2$ by flowing suspended Al/Al$_2$O$_3$/Ge core-shell NWs across the substrate. A stream of N$_2$ was pointed towards the substrate surface while simultaneously depositing a suspension of SG NWs one drop at a time. The resulting fluid flow across the surface was unidirectional and aligned the GeNWs into approximately parallel arrays. After flow deposition, the remaining fabrication steps were identical to those of the single connection SG GeNW FET with the exception of wider S/D electrodes (100μm) to afford higher number of connections as shown in Fig. 4b. While most wires lie roughly parallel to each other, variation in the orientation of the wires still resulted in some NWs crossing each other (Fig.4b inset). SG GeNW FETs with various numbers of wires up to 50 were fabricated this way. The $I_{ds}$-$V_{gs}$ curves of a FET with 35 SG GeNW connections (Fig. 4c) show an $I_{on}/I_{off}$ ~$10^4$ for $V_{ds}$ up to -1V and SS ~ 300mV/decade. The on-currents of such devices reach ~110μA (Fig. 4d) at $V_{ds}$ = -2V, consistent with the on-current of individual SG GeNW FETs. Despite of crossing of the wires, the SG scheme prevents shielding effects since each wire has its own gate stack in close proximity with the NW core. This scheme could be extended to fabricate high performance devices with SG NWs packed in three dimensions. In devices we fabricated with top-gated FETs comprised of multiple GeNWs without a surround gate, we found the on/off ratio is generally worse due to occasional crossing, stacking and thus electrostatic screening of wires.

In summary, we have demonstrated fabrication of single and multiple connection SG GeNW FETs. Our method is relatively simple and can be generalized to various semiconductor NWs to form self-aligned SG FETs on various substrates with low thermal budget. SG devices with ohmic contacts and epitaxially deposited Si shell on GeNWs are expected to afford optimum NW FETs in the future. The SG NW concept should enable new type of devices by packing SG NWs densely both in the substrate plane and into a three dimensional stack.

**Acknowledgement.** This work was supported by a SRC-AMD project, a DARPA 3D program, a NSF Graduate Research Fellowship (R.T.) and the Stanford INMP program.

**Figure Captions**

**Fig.1** Core-shell nanowires. (a) A scanning electron microscopy (SEM) image of GeNWs as-grown on a $SiO_2$ substrate with densely deposited ~ 20nm Au seed particles. The average diameter of GeNWs synthesized in the current work was ~20nm. (b) Schematic and TEM images of GeNWs after various processing steps: (i) nitride and silicon interlayer passivation by CVD, followed by (ii) atomic layer deposition of ~4nm $Al_2O_3$ and then (iii) isotropic sputter deposition of ~15nm Al. These steps led to core-shell $Al/Al_2O_3/Ge$ nanowires with a thin nitride and Si passivation layer between $Al_2O_3$ and Ge.

**Fig.2** Surround gate nanowire transistor with self-aligned source/drain and gate. (a) Schematic cross sectional views of the key fabrication steps: (i) opening of PMMA windows over the source and drain contact areas of a core-shell nanowire; (ii) KOH etching to remove Al and $Al_2O_3$ shells in the contact regions (notice undercutting in the outer shells); (iii) directional Ti deposition in source and drain regions, lift-off, followed by patterning of Pt gate electrode for contacting the surround gate. The source/drain are self-aligned with the SG shell and electrically isolated from the gate shell by the undercutting. (b) A schematic top view of the surround gate device. (c) An SEM image of a surround gate device. The surround gate (SG) metal shell is contacted by the Pt gate line (in the middle) and extends to the edges of the S/D electrodes. The inset shows a zoom-in of the drain edge next to which thinning of the SG wire (due to undercutting) is seen. Scale bar in inset is 200 nm.

**Fig.3** Electrical characteristics of a typical SG GeNW FET. (a) A 3-D schematic presentation of the device. (b) Transfer characteristics $I_{ds}$-$V_{gs}$ at various biases. The inset shows a double gate sweep of $I_{ds}$-$V_{gs}$ at $V_{ds}$ = -0.1V without any hysteresis. (c) Current-voltage characteristics $I_{ds}$-$V_{ds}$ at various gate voltages.

**Fig.4** A transistor comprised of multiple surround-gate nanowires in parallel. (a) An idealized schematic presentation of a device. (b) SEM image of a device with ~ 35 SG nanowires in parallel. Crossing wires (each with its own gate shell) are seen in the zoomed-in image (scale bar = 1μm). (c) and (d) are transfer and $I_{ds}$-$V_{ds}$ characteristics of the device respectively.

**References:**

(1)     Lauhon, L. J.; Gudiksen, M. S.; Wang, D.; Lieber, C. M. *Nature* **2002,** 420, 57-61.

(2)     Wang, D.; Dai, H. *Angew. Chemie. Int. Ed.* **2002,** 41, 4783-4786.

(3)     Wang, D.; Tu, R.; Zhang, L.; Dai, H. *Angew. Chem. Int. Ed.* **2005,** 44, 2-5.

(4)     Greytak, A. B.; Lauhon, L. J.; Gudiksen, M. S.; Lieber, C. M. *Appl. Phys. Lett.* **2004,** 84, 4176.

(5)     Xiang, J.; Lu, W.; Hu, Y.; Wu, Y.; Yan, H.; Lieber, C. M. *Nature* **2006,** 441, 489-493.

(6)     Schricker, A. D.; Joshi, S. V.; Hanrath, T.; Banerjee, S. K.; Korgel, B. A. *J. Phys. Chem. B* **2006,** 110, 6816-6823.

(7)     Kamins, T. I.; Li, X.; Williams, R. S.; Liu, X. *Nano Lett.* **2004,** 4, 503-506.

(8)     Wang, D.; Wang, Q.; Javey, A.; Tu, R.; Dai, H. *Appl. Phys. Lett.* **2003,** 83, (12), 2432-2434.

(9)     Adhikari, H.; Marshall, A. F.; Chidsey, C. E. D.; McIntyre, P. C. *Nano Lett.* **2006,** 6, 318-323.

(10)    Lu, W.; Xiang, J.; Timko, B. P.; Wu, Y.; Lieber, C. M. *PNAS* **2005,** 102, 10046-10051.

(11)    Wang, J.; Polizzi, E.; Lundstrom, M. *J. Appl. Phys.* **2004,** 96, 2192-2203.

(12)    Ng, H. T.; Han, J.; Yamada, T.; Nguyen, P.; Chen, Y. P.; Meyyappan, M. *Nano Lett.* **2004,** 4, 1247-1252.

(13)    Bryllert, T.; Wermersson, L.-E.; Froberg, L. E.; Samuelson, L. *IEEE Elec. Dev. Lett.* **2006,** 27, 323-325.

(14)    Schmidt, V.; Riel, H.; Senz, S.; Karg, S.; Riess, W.; Gosele, U. *Small* **2006,** 2, 85-88.

(15)    Goldberger, J.; Hochbaum, A. I.; Fan, R.; Yang, P. *Nano Lett.* **2006,** 6, 973-977.

(16)    Duan, X.; Niu, C.; Sahi, V.; Chen, J.; Parce, J. W.; Empedocles, S.; Goldman, J. L. *Nature* **2003,** 425, (18), 274-278.

(17)    Jin, S.; Whang, D.; McAlpine, M. C.; Friedman, R. S.; Wu, Y.; Lieber, C. M. *Nano Lett.* **2004,** 4, 915-919.

(18)    Wu, N.; Zhang, Q.; Zhu, C.; Yeo, C. C.; Whang, S. J.; Chan, D. S. H.; Li, M. F.; Cho, B. J.; Chin, A.; Kwong, D.-L.; A. Y. Du, C. H. T.; Balasubramanian, N. *Appl. Phys.Lett.* **2004,** 84, (19), 3741-3743.

(19)    Wu, N.; Zhang, Q.; Zhu, C.; Chan, D. S. H.; Li, M. F.; Balasubramanian, N.; Chin, A.; Kwong, D.-L. *Appl. Phys.Lett.* **2004,** 85, (18), 4127-4129.

(20)    Gusev, E. P.; Shang, H.; Copel, M.; Gribelyuk, M.; D'Emic, C.; Kozlowski, P.; Zabel, T. *Appl. Phys. Lett.* **2004,** 85, (12), 2334-2336.

(21)    Hausmann, D. M.; Kim, E.; Becker, J.; Gordon, R. G. *Chem. Mater.* **2002,** 4350-4358.

(22)    Groner, M. D.; Fabreguette, F. H.; Elam, J. W.; George, S. M. *Chem. Mater.* **2004,** 16, 639-645.

(23)    Williams, K. R.; Gupta, K.; Wasilik, M. *J. Microelectromech. Syst* **2003,** 12, (6), 761-778.

(24)    Tabet, N.; Al-Sadah, J.; Salim, M. *Surf. Rev. Lett.* **1999,** 6, 1053-1060.

(25)    Wang, D.; Chang, Y.-L.; Wang, Q.; Cao, J.; Farmer, D. B.; Gordon, R. G.; Dai, H. *J. Am. Chem. Soc.* **2004,** 126, 11602-11611.

(26)    Pierret, R. F., *Semiconductor Device Fundamentals*. Addison-Wesley Publishing Company: 1996.

(27)    Tu, R.; Zhang, L.; Nishi, Y.; Dai, H. *"Capacitance-Voltage Measurements of Individual Germanium Nanowire Field Effect Transistors".* To be submitted.

(28)    Krishnamohan, T.; Krivokapic, Z.; Saraswat, K. C. *IEEE International Conference on Simulation of Semiconductor Processes and Devices* **2003**, 243-246.

**(a)**

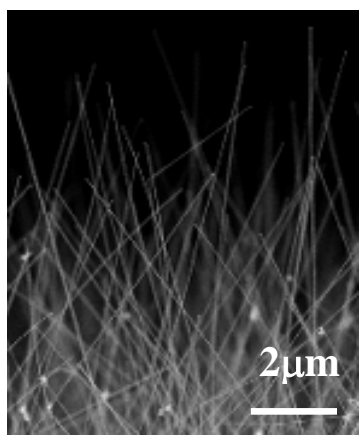

**(b)**

Si

Al$_2$O$_3$

Al

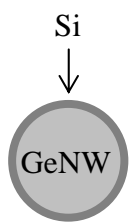

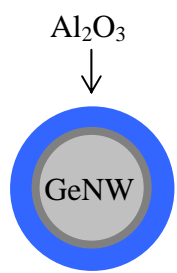

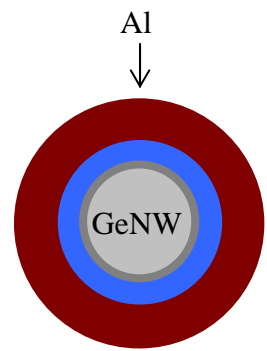

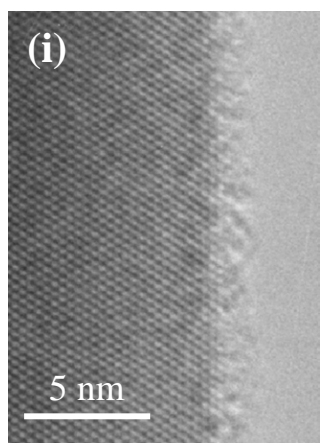

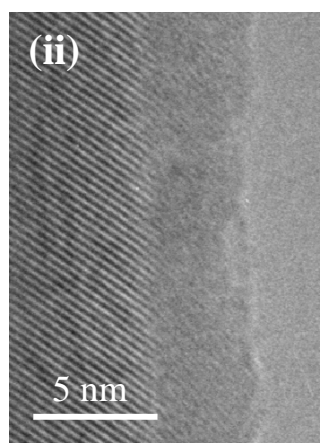

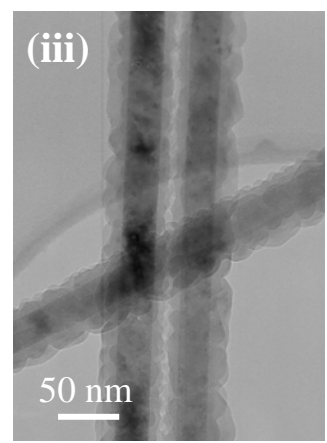

**(a)**

**(i)**

Al₂O₃ →
GeNW →
Al →

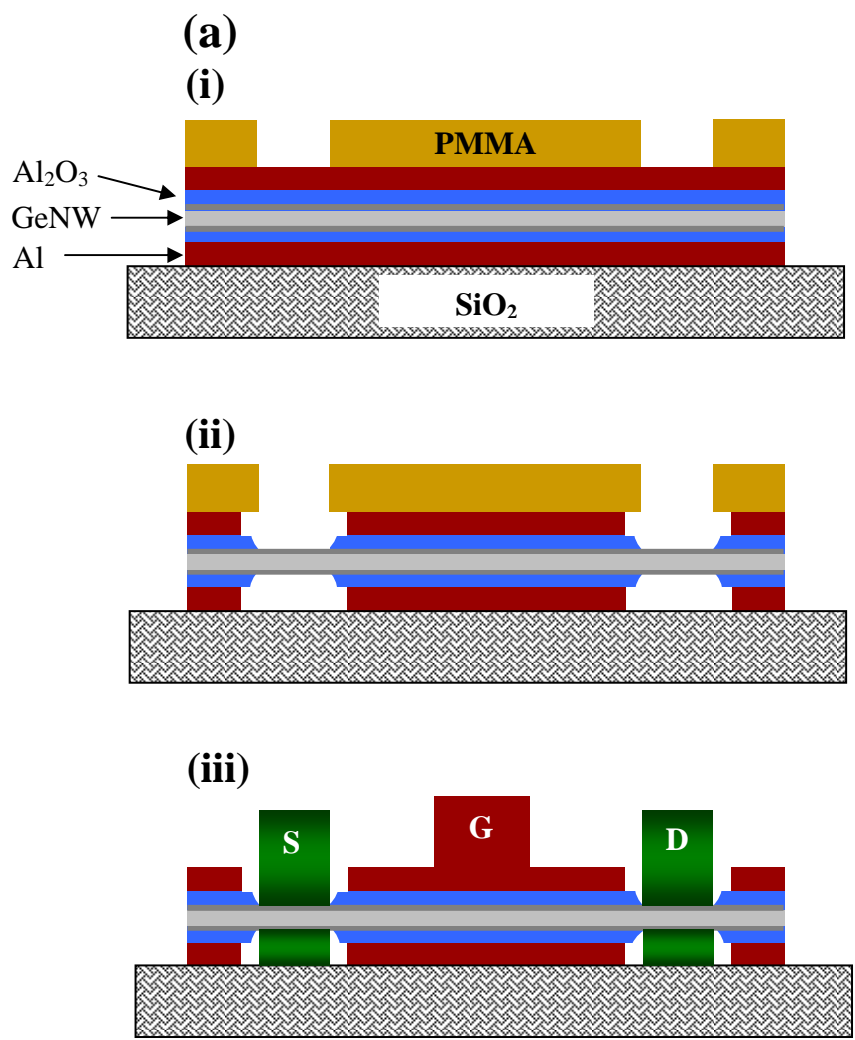

**(b)**

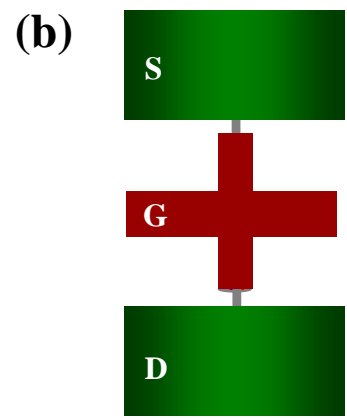

**(c)**

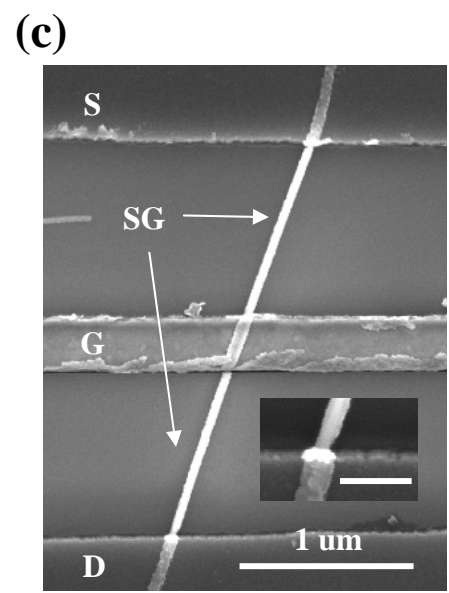

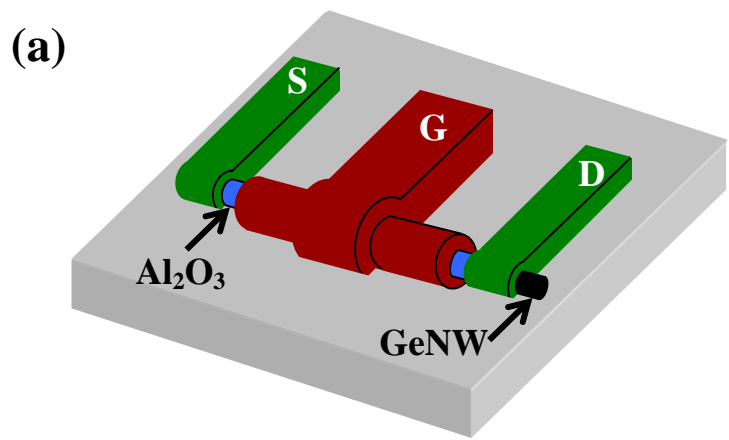

**(a)**

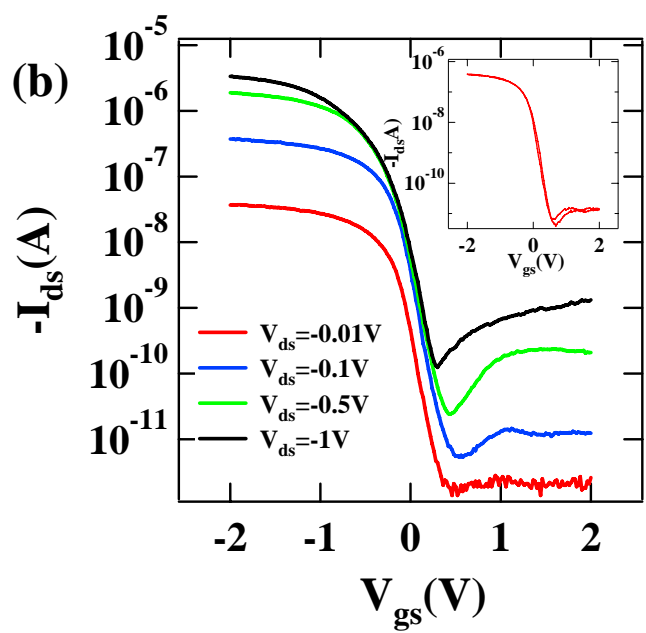

**(b)**

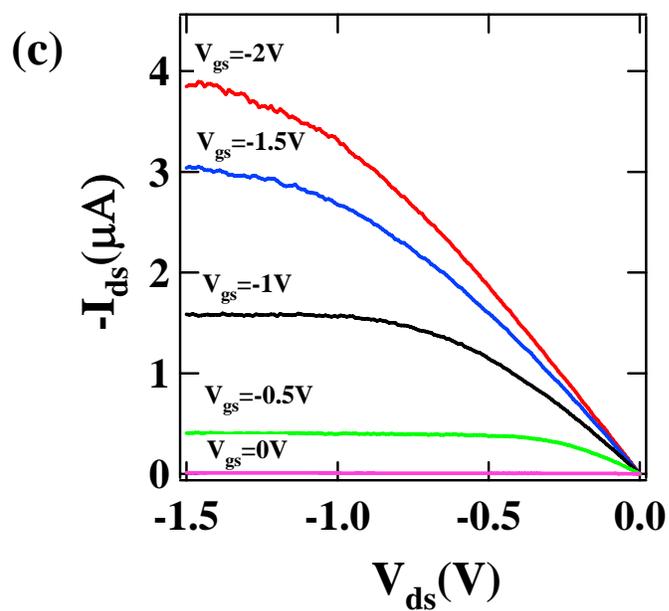

**(c)**

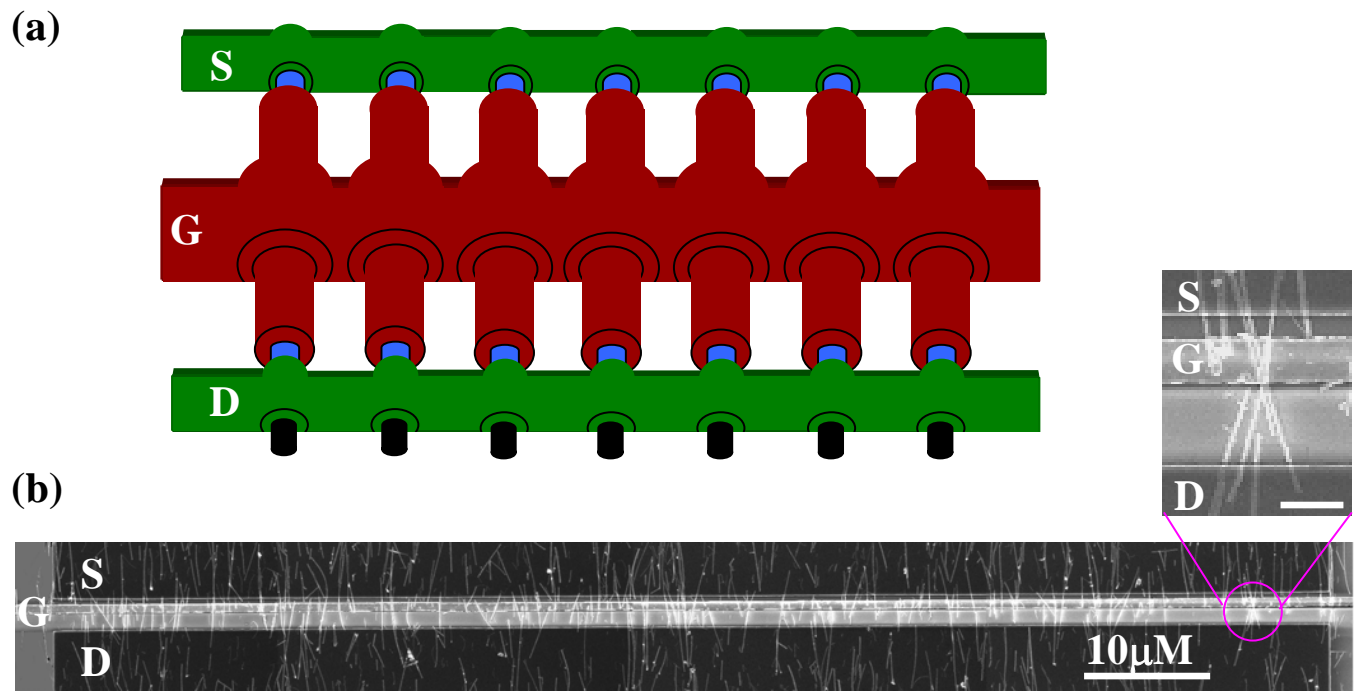

**(a)**

**(b)**

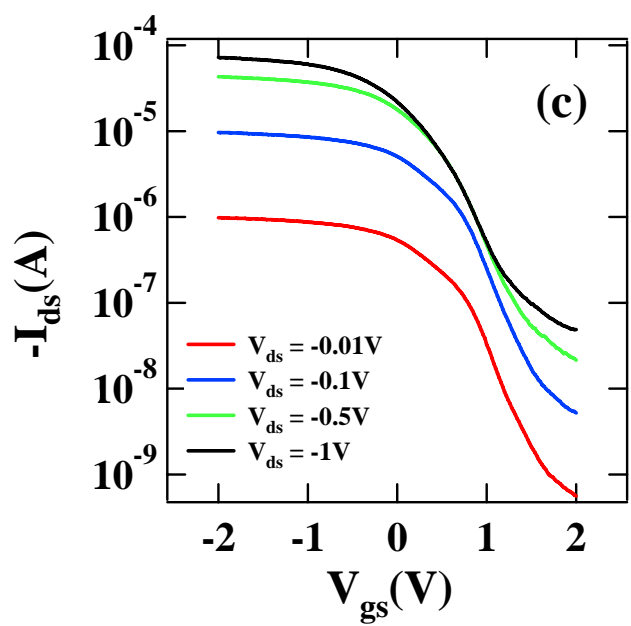

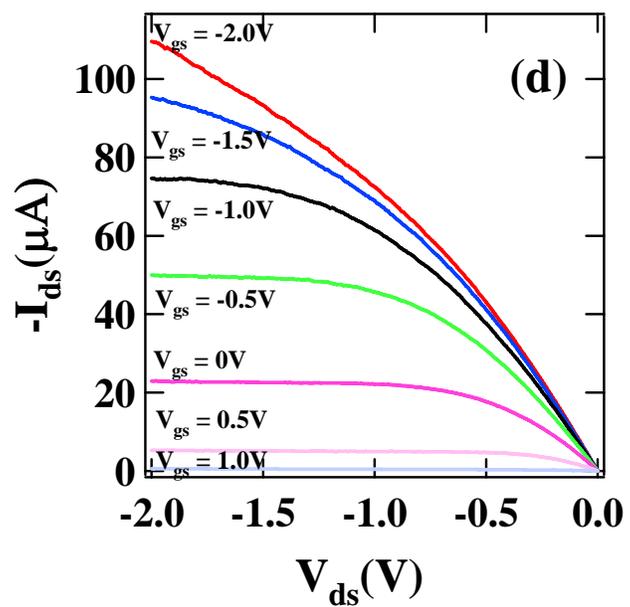